\documentclass[manuscript,screen, authorversion,nonacm]{acmart} %for arXiv

\setcopyright{acmlicensed}
\copyrightyear{2025}
\acmYear{2025}
\acmDOI{XXXXXXX.XXXXXXX}
\acmConference[MobileHCI '25]{ACM International Conference on Mobile Human-Computer Interaction}{22 September--26 September, 2025}{Sharm El-Sheikh, Egypt}
\acmISBN{978-1-4503-XXXX-X/18/06}

\usepackage{dcolumn}
\newcolumntype{d}{D{.}{.}{2}}

\begin{document}

%% The "title" command has an optional parameter,
%% allowing the author to define a "short title" to be used in page headers.
\title{Speejis: Enhancing User Experience of Mobile Voice Messaging with Automatic Visual Speech Emotion Cues}

%%
%% The "author" command and its associated commands are used to define
%% the authors and their affiliations.
%% Of note is the shared affiliation of the first two authors, and the
%% "authornote" and "authornotemark" commands
%% used to denote shared contribution to the research.
\author{Ilhan Aslan}
\email{ilas@cs.aau.dk}
\orcid{0000-0002-4803-1290}
\affiliation{%
  \institution{Aalborg University}
  \city{Aalborg}
  \country{Denmark}
}

\author{Carla F. Griggio}
\email{cfg@cs.aau.dk}
\orcid{0000-0001-9133-3828}
\affiliation{%
  \institution{Aalborg University}
  \city{Copenhagen}
  \country{Denmark}
}

\author{Henning Pohl}
\email{henning@cs.aau.dk}
\orcid{0000-0002-1420-4309}
\affiliation{%
  \institution{Aalborg University}
  \city{Aalborg}
  \country{Denmark}
}

\author{Timothy Merritt}
\email{merritt@cs.aau.dk}
\orcid{0000-0002-7851-7339}
\affiliation{%
  \institution{Aalborg University}
  \city{Aalborg}
  \country{Denmark}
}

\author{Niels van Berkel}
\email{nielsvanberkel@cs.aau.dk}
\orcid{0000-0001-5106-7692}
\affiliation{%
  \institution{Aalborg University}
  \city{Aalborg}
  \country{Denmark}
}

%%
%% By default, the full list of authors will be used in the page
%% headers. Often, this list is too long, and will overlap
%% other information printed in the page headers. This command allows
%% the author to define a more concise list
%% of authors' names for this purpose.
\renewcommand{\shortauthors}{Aslan et al.}

%%
%% The abstract is a short summary of the work to be presented in the
%% article.
\begin{abstract}
%150  or less words
Mobile messaging apps offer an increasing range of emotional expressions, such as emojis to help users manually augment their texting experiences. Accessibility of such augmentations is limited in voice messaging. With the term \emph{``speejis''} we refer to accessible emojis and other visual speech emotion cues that are created automatically from speech input alone. The paper presents an implementation of speejis and reports on a user study (\textit{N}~=~12) comparing the UX of voice messaging with and without speejis. Results show significant differences in measures such as attractiveness and stimulation and a clear preference of all participants for messaging with speejis. We highlight the benefits of using paralinguistic speech processing and continuous emotion models to enable finer grained augmentations of emotion changes and transitions within a single message in addition to augmentations of the overall tone of the message. 
\end{abstract}

%%
%% The code below is generated by the tool at http://dl.acm.org/ccs.cfm.
%% Please copy and paste the code instead of the example below.
%%
\begin{CCSXML}
<ccs2012>
   <concept>
       <concept_id>10003120</concept_id>
       <concept_desc>Human-centered computing</concept_desc>
       <concept_significance>500</concept_significance>
       </concept>
   <concept>
       <concept_id>10003120.10003123</concept_id>
       <concept_desc>Human-centered computing~Interaction design</concept_desc>
       <concept_significance>500</concept_significance>
       </concept>
   <concept>
       <concept_id>10003120.10003138.10003140</concept_id>
       <concept_desc>Human-centered computing~Ubiquitous and mobile computing systems and tools</concept_desc>
       <concept_significance>500</concept_significance>
       </concept>
 </ccs2012>
\end{CCSXML}

\ccsdesc[500]{Human-centered computing}
\ccsdesc[500]{Human-centered computing~Interaction design}
\ccsdesc[500]{Human-centered computing~Ubiquitous and mobile computing systems and tools}

%%
%% Keywords. The author(s) should pick words that accurately describe
%% the work being presented. Separate the keywords with commas.
\keywords{Affective Computing, Speech Emotion Recognition, Emotion Visualization, User Experience, Emoji, Speech-to-text, Voice message}

\begin{teaserfigure}
  \includegraphics[width=\textwidth]{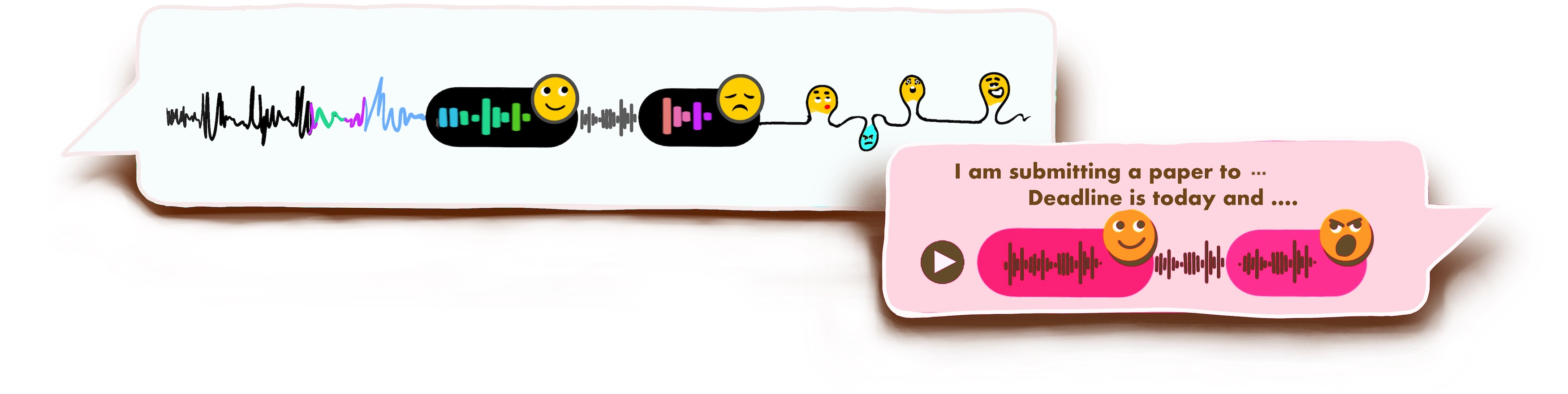}
  \caption{Illustration of the basic idea and challenges behind the concept of speejis to augment voice messaging with the help of speech emotion recognition. Users record their voice message, which is then processed for emotion recognition and ``appropriate'' speejis are placed on the waveform to give an initial impression of the sentiment of the message but also highlight emotion trajectories.}
  \Description{}
  \label{fig:teaser}
\end{teaserfigure}

%%
%% This command processes the author and affiliation and title
%% information and builds the first part of the formatted document.
\maketitle

\section{Introduction} 

One of the main use cases of mobile devices is interpersonal communication.
Here, communication options include traditional phone calls, but also text messaging, social media, location sharing, and video.
Speech plays a large role in many of these channels, be it to record a voice message, dictate a text, or instruct a voice agent.
Through speech, users can provide input while their hands are occupied, when they are on the go, but also when they need the extra expression afforded by speech over text. We can distinguish synchronous speech conversations in mobile calls from the asynchronous voice messaging, which users often fit in with other tasks and while travelling. 
Expressing oneself in digital communication often remains a cumbersome task that requires complex interactions in comparison with real-world communication.
For example, text alone often does not suffice and users instead add emojis, stickers, and GIFs to better express their emotions, moods, and intends.
Yet, adding these requires extra effort when composing the message.
This limitation extends to speech input where, even when dictation works well, the tone of the voice is not reflected to the recipient of a message.
While voice messages capture tone of voice well, this in turn requires additional effort by message recipients.
Recipients have to listen to a voice message in full, while playing these can be inappropriate in many public settings. 

With the advent of AI-based speech processing technologies, we seek to assist users in their asynchronous voice messaging and help, for example, prioritize and set expectations for incoming and outgoing messages.  
Current developments in AI for speech processing, especially speech emotion recognition (SER), achieve more accurate and more easily available speech emotion labels~\cite{wagner2023dawn, triantafyllopoulos2024interspeech}. Consequently, mobile and conversational interface designs can now exploit such technologies. Recently, An et al.~\cite{an2024emowear} proposed to use such technology to help voice message senders on a smartwatch select emotional teasers; i.e. chat-balloon animations~\cite{an2025aniballoons}, which on the receiving side can serve as enjoyable and expressive previews to an incoming voice message, allowing a glimpse to their potential context. 
Beyond smartwatch users, we believe that users messaging from smartphones and personal computers could benefit from more detailed previews of voice message contents, e.g., a transcription alongside emotional cues.

The mixed design space of affective speech AI~\cite{captionroyale} and voice messaging~\cite{weber2023} presents unexplored opportunities to improve the user experience and accessibility of modern messaging practices. 
In particular, we are interested in exploring opportunities related to deeper integrations of SER to help improve design and delivery of augmentations with the use of continuos speech emotion models opposed to emotion categories. 
While previous work analysed the emotion category and visualized the emotional tone of an entire voice message~\cite{an2024emowear}, we are also interested in exploring the value of segmenting the analysis of messages in multiple parts to convey the varying emotional cues throughout a message, which may be particularly convenient for longer messages.  
Overall, we aim to contribute to this larger endeavour by exploring finer-grained augmentations of voice messages to discover the opportunities and limitations of this design space.

To this end, this paper introduces `speejis'; emojis and other visual speech emotion cues that are created automatically from speech input alone. Ultimately, the speejis system is an implementation of a proactive agent and a realization of an implicit interaction design where voice message augmentations are created based on the tone of the sender. However, speejis can also be used more explicitly, whereas the user can exaggerate the tone of their voice or add non verbal emotional sounds, such as laughter or ``hmmm'' for more direct control of the speeji selection. 
To study the use and effect of speejis on user experience, we present a functional prototype and report on a user study with speeji design probes. Our results show that speejis are generally perceived as desirable voice message augmentations, including a high ease of use and an increase of expressiveness in emotional communication. We outline design implications and highlight future research directions.   

\section{Background}

With computing becoming increasingly ubiquitous and mobile over the last two decades AI-based and inspired solutions played their part in enabling users to address the limitations and potential of using a small device on the go. 
Key to the success and the  popularity of mobile applications has been the use of context through multiple sensors~\cite{gellersen2002multi} and lenses~\cite{raptis2005context}, and adapting interaction techniques~\cite{aslan2005compass2008} and information presentation~\cite{kruger2004connected} to the needs of mobile device users.
The computational limitations of mobile devices have heavily shaped how user interfaces needed to be designed~\cite{jones2006mobile} with researchers in mobile HCI exploring alternative techniques to address computation limitations and extend interaction spaces by, for example, using multiple devices at once and integrating additional displays (e.g.,~\cite{grubert2015multifi}). Technology and techniques to recognize context and human activities through mobiles have evolved~\cite{Yin24_survey} with deep learning approaches improving in robustness and as deployable options on mobiles~\cite{Kunwar_2022}.
Overall mobile interfaces are becoming increasingly ``intelligent''. Machine learning technologies play an increasing role in this development, helping,  for example, to  distinguish between different touches~\cite{Harrison2011}, recognizing hand gestures from ultrasound data~\cite{McIntosh2017}, tracking micro-gestures~\cite{Wang2016}, or enabling speech command recognition~\cite{juang2005automatic}. Furthermore, models can be used to augment existing interaction modalities by as aforementioned classifying contexts in mobile computing~\cite{Krause2006}, detecting social situations~\cite{Tan2022}, or assessing user's interruptibility~\cite{Zuger2018}.

With natural language processing and language models becoming increasingly accessible, speech input is once again a driver for innovations. Sharon Oviatt~\cite{oviatt_1996}  highlights that spoken, written, keyboard, and other modalities shape the language transmitted within them. Today, we are witnessing how the recent progress in (generative) AI is impacting content creation and consumption. Mobiles are progressing ever more into new and converged media or medium. This is happening in the basic sense proposed by McLuhan~\cite{mcluhan1967medium} and his attention on sensory effects, society, and transitions from one medium to the next, but also arguing in a similar fashion to modalities that the medium itself becomes part of the information. Experience design in conversational user interaction on mobiles and mobile mediated communication is expected to be further shaped by interaction modality and AI integration. We believe that some inspiration and benefit can be taken from related fields, especially the field of human-robot interaction where previous work has, for example, exploited laughter and smile recognition models for implicit~\cite{Ju_2015} and adaptive interaction designs, allowing robots to adapt to the humour of their human conversational partners~\cite{weber2018shape}, or for a smart mirror to proactively generate personalized compliments based on recognizing visible features of the specific user~\cite{bittner2019smarthomes}. 

\subsection{Mobile Messaging and Accessibility}

Messaging platforms have historically been designed for text-based communication, a medium that inherently lacks support for non-verbal expression (e.g., a smile). 
Users have leveraged this limitation as a new opportunity to control the impression they convey to others, for example, by adding a smiley emoticon :) to a message even when they are not actually smiling~\cite{walther2001impacts,WALTHER20072538}. 
However, the lack of non-verbal and intonation cues can often lead to misinterpreting the intended tone or meaning of a message.
This has motivated the creation and adoption of diverse visual means of non-verbal expression in messaging platforms, such as GIFs~\cite{jiang2018}, stickers~\cite{cha2018,zhou2017} and emojis, which have been adopted for diverse expressive functions, such as expressing emotions and clarifying the tone of a message~\cite{cramer2016}. 
Recent research highlights a long-standing interest in augmenting text-based communication with richer, non-verbal means of expression beyond GIFs, stickers and emoji~\cite{Buschek_2018}, for example, by allowing manual or automated customizations to the shape~\cite{emoballoon} and color~\cite{heartchat} of message bubbles, keyboard color themes~\cite{dearboard}, or animations~\cite{mojiboard} and vibration patterns associated to an emoji~\cite{vibmoji}.

In recent years, voice messages have seen widespread adoption among users of messaging platforms, even though the functionality has been available in some mainstream apps (e.g., WhatsApp, WeChat) for over a decade. 
Voice messages, in contrast to text messages, support much richer expression. Moreover, unlike phone calls, they benefit from the asynchronous nature of messaging.
The increasing popularity of voice messages starts to uncover challenges and opportunities characteristic of mixing text-based and voice-based messages in the same communication channel: 
voice messages are convenient for sending long messages, however, listening to them requires increased effort~\cite{Haas2020b}. 
Moreover, the mix of text and voice messages both reduces and causes \textit{situational impairments}~\cite{Wobbrock2006TheFO,Sarsenbayeva17} inherent in mobile messaging contexts. 
For example, listening to a voice message or recording a voice message may be preferred while walking~\cite{walktype,Vadas2006} or when messaging in a cold environment~\cite{Sarsenbayeva17} instead of holding the phone to read and type. 
Conversely, listening to a voice message may be challenging or inappropriate in some situations, such as when messaging in very loud environments (e.g., a subway, a party) or in the presence of others (e.g., in a meeting)~\cite{weber2023}. 

To address these challenges, research has explored how to bridge the accessibility gap between voice and text modalities.
%no control of expression/intonation in text-to-speech:
For example, when bridging text messages to speech interfaces (e.g., screen readers or voice assistants), senders may use text-based expressions that result in awkward or absent voiced expressions (e.g., emoji with unexpected descriptors or exaggerated exclamation marks), for which researchers have recommended accessibility guidelines~\cite{griggio2024, tigwell2020}. 
%transcriptions don't show intonation/tone/expression
In the case of speech-to-text interfaces, such as automatic transcriptions of voice messages, non-verbal expression and intonation nuances also get lost in the change of modality, resulting in a plain transcription with no cues about the message's intended tone.
To explore how to convey tone and emotional expression in speech-to-text interfaces, researchers have visualized the emotional tone of a voice message with special ``message bubble'' designs: AniBalloons and EmoWear map diverse animations on the voice message bubble to reflect the emotions expressed in the message content~\cite{an2024emowear,an2025aniballoons}, and Chen et al. map emotions to different bubble colors~\cite{bubblecoloring}.
VoiceMessage++~\cite{Haas_2020_mobileHCI} enables sound augmentations for voice messages (e.g. background music or a one-shot sound effect such as a party horn) and visualizes these in the form of emojis decorating the waveform of the voice message.
This paper extends this work by contributing a novel design for conveying tone and emotional expression in speech-to-text interfaces alongside transcriptions.

%motivate the need to bridge voice messages to a text-based medium from a situational accessibility perspective: look for papers motivating speech to text interfaces. 

%challenges related to cross-modality messaging. More and more, conversations are a mix of text and voice messages
    %voice messages are embedded in a medium primarily dedicated to asynchronous text-based conversations
    %we focus on speech to text challenges to bridge the gap between voice messages and a visual, text-based conversation
        %aniballoon, emowear, emoballoon, what else?

%why basing speejis on emojis?
%non-verbal cues in messaging: 
    %emoji, evidence of adoption for expressing emotions and tone (Cramer, etc) 
    %it's something that users already know how to interpret so we can rely on their previous knowledge/adoption (Griggio 2019? %voicemoji?)
    %other efforts in augmenting expression
        %vibmoji
        %buschek's fonts
        %sally's animoji
        %dearboard

%for implications/discussion: potential for customizing the speejis scale of valence/arousal based on relationship-specific interpretations of emoji (Griggio 2019, Wiseman, Tigwell/Miller, Kelly, Dearboard...)

\subsection{Speech Emotion Recognition (SER)}
Within the area of affective computing~\cite{picard2000affective}, an emerging aspect is speech emotion recognition (SER). Recent deep learning approaches have improved many tasks within affective computing~\cite{Triantafyllopoulos_et_al_2024}, including automatic analysis of speech and identification of the emotions it conveys~\cite{Wani_et_al_2021}. 

Emotions in speech can be recognized by analysing both linguistic and paralinguistic features. Both types of emotion recognition describing \textit{how} something is said and \textit{what} is said have their strengths and limitations. While humans can clearly voice fine-grained semantics with verbal utterances combining words and sentences, acoustic features and non-verbal sounds, which can be short and effective, have a way of providing emotional context to what is being said. This can of course be used to complement and multiply an emotion that is already carried in the verbal content. But it can also be used to contradict the verbal content producing ironical utterances, which have their own place and function in social interactions, including human interactions with social robots ~\cite{ritschel2019irony}. 
Therefore, speech emotion recognition based on paralinguistic features is arguably a better fit for augmenting voice messages, especially when transcriptions can be used to provide an additional (pre)view of what is said in a voice message to improve robustness.
In fact, if one would want to use SER based on linguistic features, they would have to perform automatic speech recognition (ASR) first in order to transcribe speech into text, which of course delays the speech emotion processing significantly. On mobiles ASR is something that is typically moved to the cloud while SER based on paralinguistic features could achieved on the mobile, providing an additional degree of privacy. It is also known that the accuracy of ASR suffers from various factors, such as background noise or speaker dialect. Moreover, word (and character) errors impact the performance of the subsequent speech emotion recognition task based on linguistic features~\cite{amiriparian2021impact}. 

Traditionally, emotion recognition tasks in machine learning (ML) focus on two types of emotion models, recognizing (i) emotion labels for discrete emotion categories such as sadness, joy, or anger, which are often referred to as basic emotions; and recognizing values in (ii) continuous emotion dimensions, such as valence and arousal. Arguably, continuous dimensions are of greater utility and flexibility. Moreover, emotion categories are of limited use if one is interested in studying (continuous) emotion transitions and trajectories~\cite{christ2024modeling}.

Considering the field and task of speech emotion recognition (SER), important progress has been achieved by using transformer architectures, successfully closing the `valence gap'~\cite{wagner2023dawn}; i.e., successfully using paralinguistic features only to recognize valence in speech emotion recognition tasks. The field of SER has benefitted from various past emotion recognition challenges, with the first official challenge happening at INTERSPEECH 2009~\cite{schuller2009interspeech}. Since then, the field has progressed through superior methods, primarily based on deep neural networks to tackle the task of recognizing emotions in speech. Today, SER is increasingly used to recognize continuous emotion dimension values, especially for the dimensions of valence, arousal, and dominance. Reflecting on past SER challenges, Triantafyllopoulos et al.~\cite{triantafyllopoulos2024interspeech} describe how SER methods have evolved throughout the last 15 years from multi-layered perceptrons, long short-term memory recurrent NNs, convolutional NNs, to transformer based models and self-supervised learning approaches.

\begin{figure}[ht!]
  \centering
  \includegraphics[width=\linewidth]{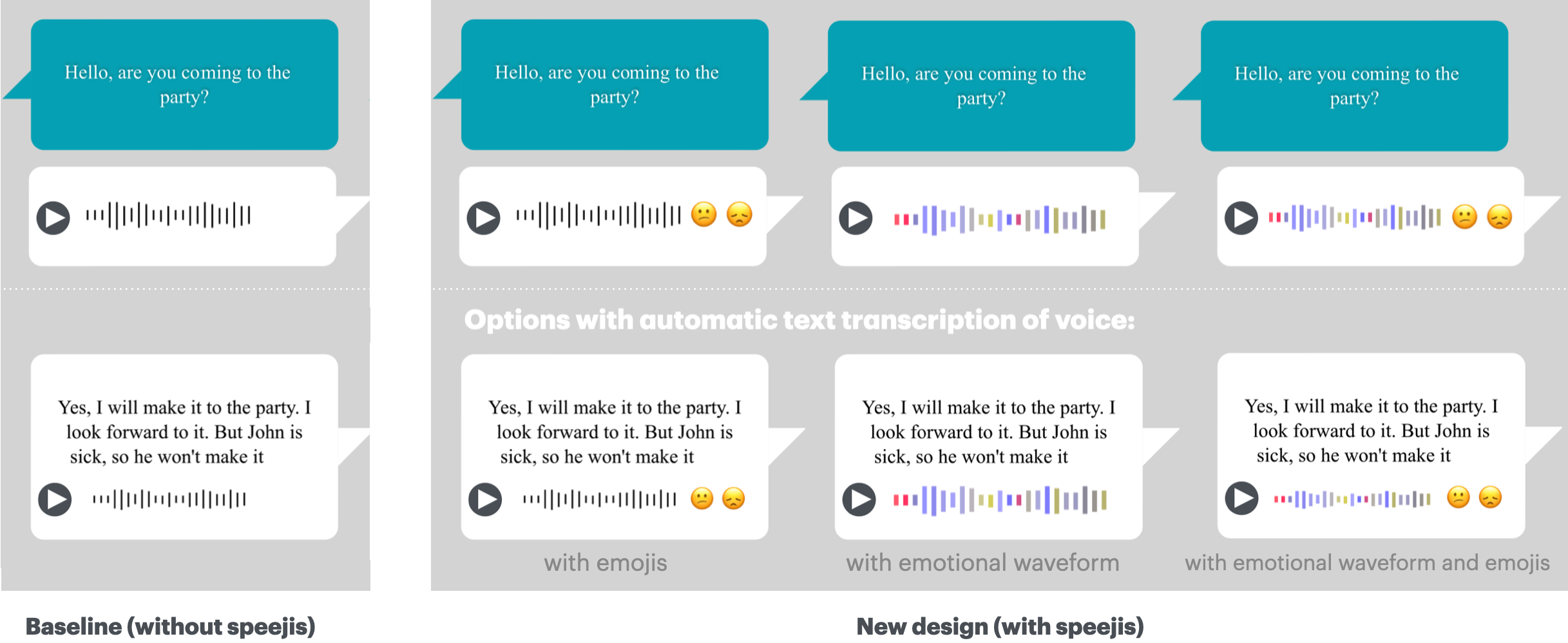}
  \caption{Overview of design probes used in the study, representing conditions for baseline voice messages without speejis (left) and the SER augmented design with speejis (right).}
  \Description{}
  \label{fig:study_conditions}
\end{figure}

\section{Speejis Prototype}
To study the effects of using SER to help design and automate the delivery of speejis in voice messages, we implemented speejis as a web application. The Flask python web framework is used as the backbone, delivering web-based user interfaces which can be customized for mobile usage. Consequently, the visual realization of speejis in our implementation is based on web front-end technology, specifically HTML emojis and CSS. Figure~\ref{fig:study_conditions} illustrates an example voice message with and without speejis.

\subsection{Speeji Designs}
\begin{figure}[ht!]
  \centering
  \includegraphics[width=\linewidth]{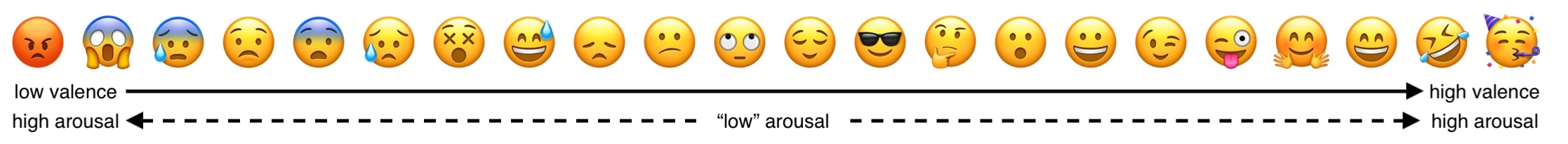}
  \caption{The 22 emojis used as speejis in the study to automatically augment voice messages.}
  \Description{}
  \label{fig:emojis}
\end{figure}

We realized two speeji designs, which we regard as starting points in the design of speejis. First of all, we aimed to deliver emojis based on SER results. To this end we make use of the emoji classifications provided by Kutsuzawa et al.~\cite{kutsuzawa2022classification}, who provide valence and arousal values for 74 facial emojis. Figure~\ref{fig:emojis} shows the 22 of these emojis which we selected for our implementation and later used in the user study, representing a wide range of possible emotions. We acknowledge that there could be limitations in using a subset of the typically available emojis in text messaging applications. However, emojis have been highly successful in augmenting text messaging and users are very accustomed to using and interpreting emojis. Therefore, we selected emojis representing a broad range of emotional expressions, from low to high valence and arousal. We argue that this diverse selection is sufficient to demonstrate a proof-of-concept system that end users can evaluate, helping to identify both its limitations and potential in terms of user experience. 

As smartphones increasingly offer text transcripts of voice messages, we have included text transcripts as part of voice messaging UIs. It should be noted that text transcripts are not yet a standard for voice messaging apps for multiple reasons, including language and accuracy--lower accuracies in automatic speech recognition could impact speech emotion recognition results~\cite{amiriparian2021impact}. In our design explorations, we added emojis to augment both the waveform and text transcripts alone or both. Figure~\ref{fig:study_conditions} shows the design probes of how our system uses emojis to augment voice messages compared to how voice messages look and feel without speejis. 

A benefit of using paralinguistic features for speech emotion recognition is that the audio stream can be split into different segment lengths of audio chunks and fed into the SER model to get emotion recognition results for all parts of the message in different granularity depending on the size of the chunks. There are, naturally, some limitations associated with choosing too small or large chunks and accurately summarizing the message sentiment. However, there are also benefits in running the SER for distinct parts as well as the overall message. Control over the chunk size could be highly beneficial for designers of augmentations, providing space for further design exploration and additional functional benefits.

Based on our own explorations and feedback from a pilot study, we considered which part of the message to augment with emojis and how many emojis to show for each message. Based on this, we created probes using two emojis---with the first emoji referring to the overall sentiment of the voice message, while the second emoji refers to the very last part of the voice message. This was motivated by the observation that the affect expression in the last part of a voice message is often more actionable and implicitly directed towards the receiver of the messages as an indication for expected communication flow. We chose not to add an emoji to the beginning of the message, which often serves the role of introduction.  

\begin{figure}[ht!]
  \centering
  \includegraphics[width=\linewidth]{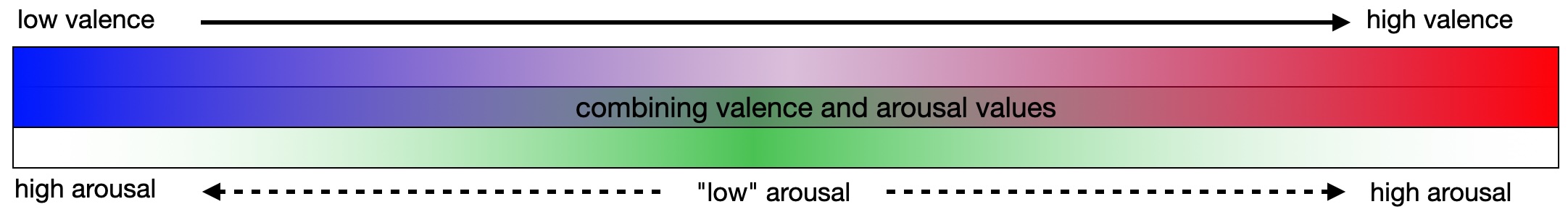}
  \caption{Concept for the colour mapping used to augment the audio waveform  and create an emotional waveform.}
  \Description{}
  \label{fig:colormapping}
\end{figure}

In addition to using emojis as symbolic augmentation, we explored ways to also augment the audio waveform, which is typically shown as part of a voice message. The standard waveform illustrates progress in amplitude of the audio over time, indicating when there is higher and lower volume of sound in the message. The waveform can also be useful during playback of the voice message allowing users to play or slide to skip to other parts of the message. 

We created a first version of an emotional audio waveform using a simple colour mapping (see Figure~\ref{fig:colormapping}) to indicate valence and arousal for individual bars in the audio waveform. To determine the smallest possible audio chunk size that maintains accuracy with the SER module, we tested how much we could reduce the size before it started affecting the reliability of the results. We settled on using audio chunks of 0.5 seconds. For the colour mapping we oriented ourselves by the mapping indicated in the graphs used by Kutsuzawa et al.~\cite{kutsuzawa2022classification}. Clearly, using colour mappings comes with some limitations and issues, including individual and cultural differences. Given the initial stage of our design explorations and studies, we decided it would be a good starting point to support the user study and obtain initial insights. 

\begin{figure}[ht!]
  \centering
  \includegraphics[width=\linewidth]{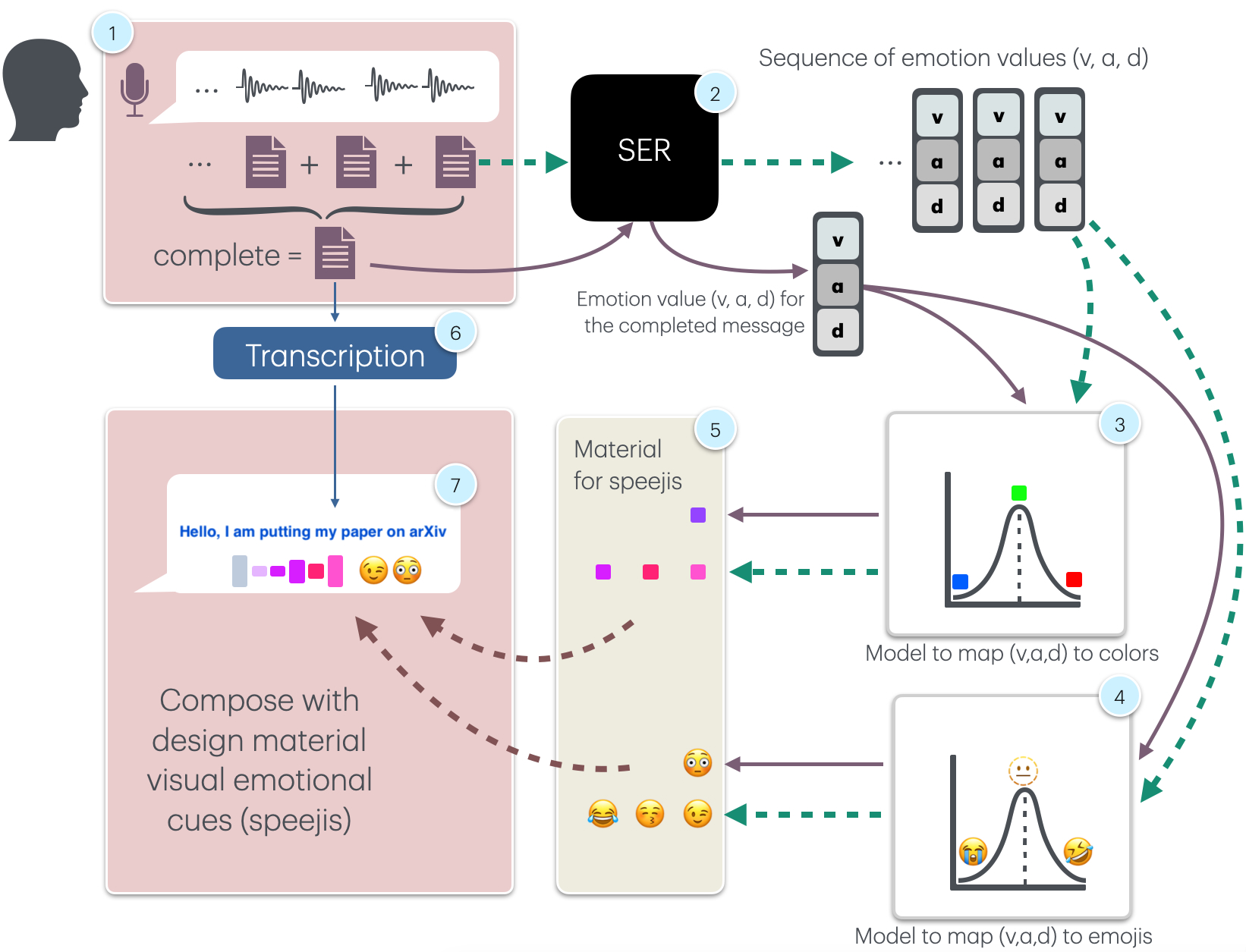}
  \caption{Overview of the speejis system illustrating the components of the system and how they interact with each other to provide the material needed to compose speejis and augment voice messaging UIs.}
  \Description{}
  \label{fig:prototype}
\end{figure}

\subsection{Speejis System Overview}

Figure~\ref{fig:prototype} shows an overview of the speejis system, which is used to process the audio stream from a microphone (\textbf{1}). A SER model (\textbf{2}) provides emotion values of valence, arousal, and dominance for both the smaller audio chunks and the whole message. 
The whole message is saved as an audio file, which is processed by a transcription model (\textbf{6}). To perform SER based on paralinguistic features, we use the model provided by Wagner et. al~\cite{wagner2023dawn}. To transcribe the voice message into text, we use the base Whisper model~\cite{radford23a}. The two aforementioned mapping models (see Figure~\ref{fig:emojis} and~\ref{fig:colormapping}) are applied to map emotion labels to emojis (\textbf{3}) and waveform colours (\textbf{4}), resulting in the design material (\textbf{5}) which can be used to compose visual cues based on emojis, colours, or both. The resulting visual cues can be added to the user interface for the voice message. 

We implemented the system as a web solution using Flask and python as back-end technology, capable of serving websites to various mobile devices (based on HTML, CSS, and JavaScript).

\section{User Study}
To evaluate the user experience of speejis in voice messages, we set up a user study to compare the UX of voice messaging with and without speejis. The goal of the study was to test our assumption that speejis are usable and that their use could improve the messaging experience. We were interested in obtaining insights from users on the potential benefits and limitations to inform and improve the design of speejis. We also sought to gather higher-level perspectives on the use of speech emotion recognition technology both as sender or receiver of speech emotion augmented voice messages. An iPhone 12 Pro was used as the front-end to experience voice messaging, while the Speejis system was run on a Macbook Air (M4). The iPhone and MacBook Air were connected through a dedicated WiFi access point to enable communication for all of the back-end processing. 
A desktop condensor microphone was used to both stream the voice message and record the posthoc interviews with the participants. We planned 30 minutes for each session and this worked well for the majority of the participants, while some took a few minutes longer when discussing potential implications of using SER in voice messaging. 

We chose not to use the iPhone microphone to process the audio to (i) bypass risks of specific preprocessing steps the iPhone may apply to handle specifics of their microphone arrays, such as using audio enhancement modules, and (ii) since the SER model that we used was trained on podcasting data, which typically uses the kind of desktop condensor microphone that we had at our disposal. However, we seated the participants next to the desktop condensor microphone and let them believe that the recording was happening on the mobile. When participants played the voice message, the audio of the recording was played on the iPhone.

\subsection{Participants}
We recruited 12 (eight female, four male), students (six) and staff members from the University campus between the ages of 25-44 years (M=29), with varying experiences in using voice messaging in their daily lives.

\begin{figure}[ht!]
  \centering
  \includegraphics[width=.8\linewidth]{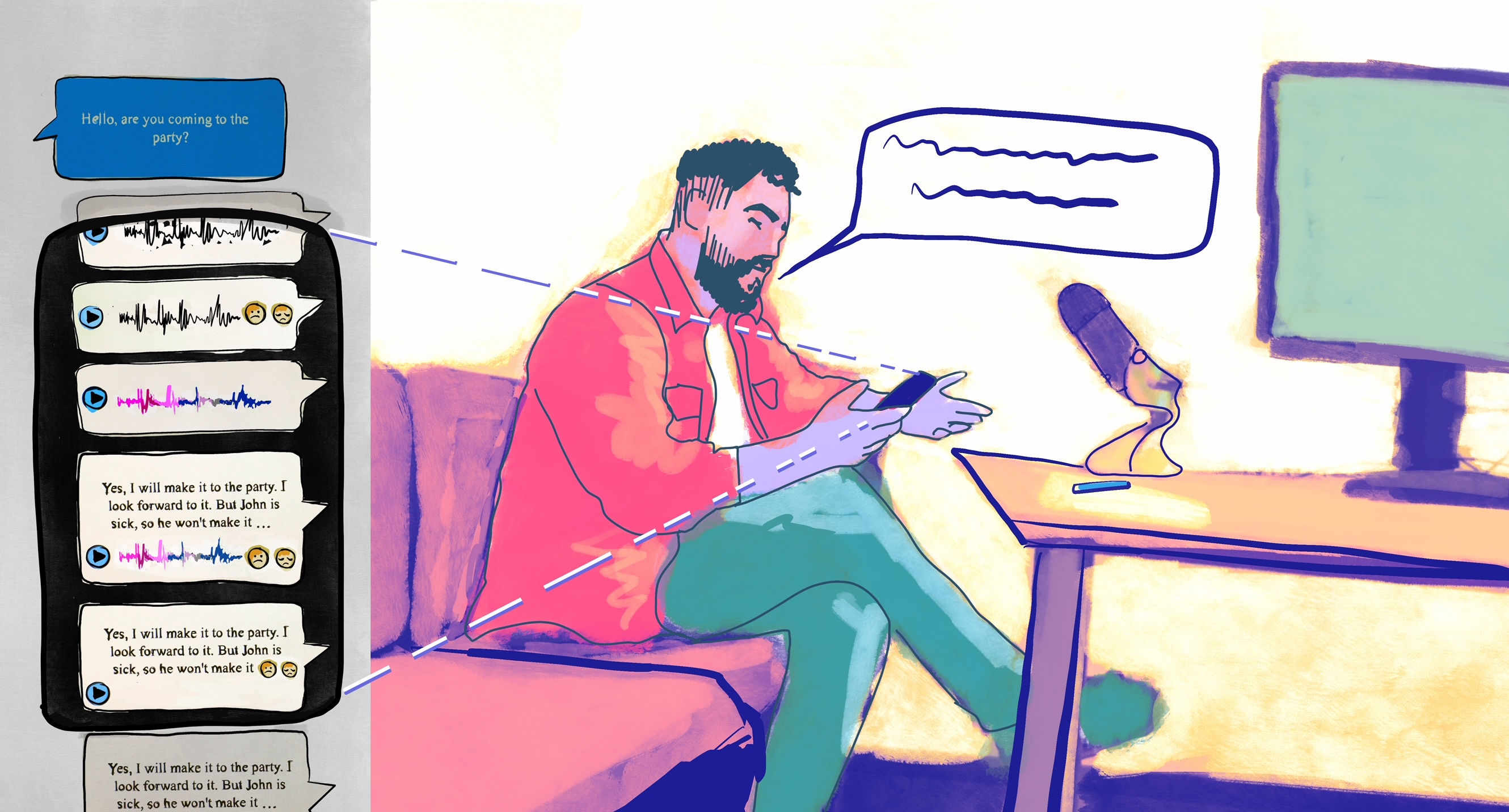}
  \caption{Illustration of study setup.}
  \Description{}
  \label{fig:setupstudy}
\end{figure}

\subsection{Procedure}
Figure~\ref{fig:setupstudy} illustrates the setting for the user study. We allowed participants to take part in the study alone or in groups of two in order to gather detailed feedback from single users and deeper discussion among pairs that know and communicate regularly with each other. Three groups of two participants took part in the study, while the rest participated individually. The number of participants in the session did not change the general procedure of welcoming the participants, explaining the basic concept of speejis, demonstrating the system, and showing them examples of voice messages with and without speejis. The participants then used the system for as long as they wanted. 

Participants were presented with the first view, which resembled a messaging UI showing a speech bubble with the message, ``Are you coming to the party?''. We instructed participants that they could leave any voice message they wanted. 
Participants could leave a voice message by using the `start recording' and `stop recording' buttons. The start button started recording and processing the user's speech data and the stop button resulted in showing the voice message of the user in all designs (see Figure \ref{fig:study_conditions}), one below the other--with and without speejis mixed in order. Participants could scroll down through all of the displayed voice message designs. Two buttons were placed after the last design where they could also listen to an audio recording of the voice message and return to test the system again with a subsequent message.  
All participants were encouraged to explore the system and try leaving messages with different intended feelings, e.g. happy, sad, or even messages with an angry tone. Participants were also invited to record mixed emotions in the message to transition the emotional tones in the same message, e.g., starting with a happy and ending on a sad emotional tone.  

When participants had finished exploring the prototype and were ready to provide us feedback, they were first asked to fill out the English version of the UEQ~\cite{Laugwitz_et_al_08} --- both for the voice messaging experience with and without speejis. 
Afterwards, we conducted a semi-structured interview during which we encouraged participants to go back to the prototype and test it again, to reflect on their comments, or demonstrate to us what they were referring to for clarification. We started the interview by asking if they would use speejis in their personal communication, followed by which version of the design they would prefer to use and why. We asked them to explain what they liked and disliked, if their answers or preferences would change if they were on the message receiving side. We also asked if they had any suggestions for improving future design iterations.

The last part of the interview focused on higher level implications e.g. if the speejis would influence conversation flow, how they felt about the level of automation, and any other topics they would want to share with us regarding how they imagined this being used in daily life.  

\section{Results}
We first present the results of the UEQ survey, which measures pragmatic and hedonic constructs, followed by the results and analysis of the semi-structured interviews.

\subsection{Analysis of the UEQ Data}

\begin{table*}[t]
\caption{Overview of study results for the UEQ.}
\label{tbl:ueqscores}
\centering
\begin{tabular}{l dd dd r dl}
\toprule
 & \multicolumn{2}{c}{With Feedback} & \multicolumn{2}{c}{Without Feedback} \\
 \cmidrule(r){2-3} \cmidrule(l){4-5}
Dimension & \multicolumn{1}{r}{Mean} & \multicolumn{1}{r}{SD} & \multicolumn{1}{r}{Mean} & \multicolumn{1}{r}{SD} & t-statistic & \multicolumn{2}{c}{p-value} \\
\midrule
Attractiveness & 1.8 & 0.8 & 0.6 & 0.7 & t(11) = 4.01 & 0.0021 & $\ast\ast$ \\
Dependability & 0.6 & 0.7 &  1.3 & 0.7 & t(11) = -3.12 & 0.0097 & $\ast\ast$ \\
Efficiency & 1.4 & 0.9 &  0.9 & 1.2 & t(11) = 1.08 & 0.303 & \\
Novelty & 1.9 & 0.6 & -0.6 & 1.5 & t(11) = 2.56 & 0.0002 & $\ast$$\ast\ast$ \\
Perspicuity & 1.5 & 1.2 &  1.5 & 1.3 & t(11) = 0.04 & 0.970 & \\
Stimulation & 1.9 & 0.6 & -0.1 & 1.5 & t(11) = 4.42 & 0.0010 & $\ast\ast$ \\
\bottomrule
\multicolumn{8}{l}{\textit{Note:} $\ast\, p < 0.05$; $\ast$$\ast\, p < 0.01$; $\ast$$\ast$$\ast\, p < 0.001$} \\
\end{tabular}
\end{table*}

\begin{figure}[tb]
  \centering
  \includegraphics[width=0.7\textwidth]{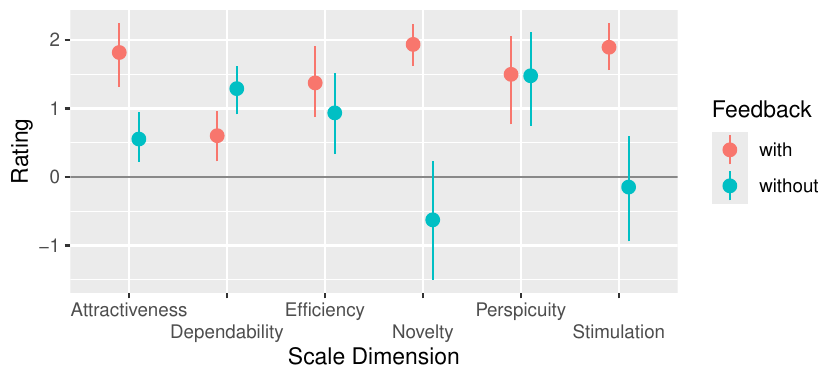}
  \caption{Results of the UX questionnaire comparing voice messaging experience with and without visual speech emotion cues. Error bars denote 95\% confidence intervals.}
  \Description{TODO}
  \label{fig:ueqscores}
\end{figure}

We ran a paired sample t-test for each dimension of the UEQ, finding significant differences for attractiveness, novelty, perspicuity, and stimulation (see Table~\ref{tbl:ueqscores} and Figure~\ref{fig:ueqscores}).
Inclusion of speech emotion cues resulted in increased attractiveness, novelty, and stimulation, but came at a slight cost to dependability.
In the following section we will try to shed some light and explain participant ratings through the analysis of the interview data.

\subsection{Analysis of the semi-structured interview}

\subsubsection{Overview of participant preferences}

All (12) participants reported that they would want to use voice messaging with speeijis and all of them also would like to have the transcription function. All participants preferred emojis to be included in the design. Eight of the 12 participants preferred the combination of emojis with coloured waveforms, arguing, for example, that it could provide the function to skip to the more sad or interesting parts of the message and that it worked well with the ending emoji which described the emotions at the ending of the message. The other four participants who preferred to use emojis with the standard waveform argued that the colours were adding too much complexity, and that the colour mappings were not how they would have expected it to be. One of these participants argued that the coloured waveform had a ``too quantifying feeling to it...'' and was also ``too jumpy and not really aesthetic''. 

\subsubsection{Detailed feedback on design features}
Participants provided very specific feedback about the emojis, text summaries, the waveform, and SER and augmentation.

Emojis were described as comforting and recognizable. Two participants mentioned that emojis have some room for ambiguity which they actually liked. There was no concrete agreement on the number of emojis and on which parts of the message emojis should represent. 

Participants highlighted that longer message may require more emojis and that maybe emojis should not be applied to any region of the message, and especially to regions where the emotionality is pretty neutral but instead emojis could only be used to explain parts of the message with extreme values. Most participants appreciated the fact that the speech emotion recognition was only using how participants sounded in the message and what they had said which was provided by the text transcript. However four participants brought up the topic of also including linguistic analysis for the augmentation and one participant suggested that it would be nice to have the content of the message being integrated into the emojis, such as when a horse riding is mentioned that emoji could integrate this information with the emotional sound. Two participants mentioned animations as an alternative to improve the coloured waveform, having for example each bar move according to its emotion labels, which would add to liveliness of the design. Two participants mentioned that text summaries (e.g., based on large language models) are getting popular but that they dislike them since they change the conversations and they would hope the receiver would listen to the voice message anyway. They added that in that case, it would not even matter much if the emotion augmentations were not perfect.

%Indeed, such emojis would make even more sense for long messages.  

\subsubsection{What did participants like?}
Participants liked that speejis were considered as fun, nice, supportive, helpful, providing an added level of communication, making it more lively, and much quicker and easier to get a sense of what is going on, to mentioned a few of the formulations participants used to describe what they liked about emojis. These comments are in line with  the results of the UEQ questionnaire where we saw that participants rated overall attractiveness, novelty, and stimulation dimensions of voice messages with speejis significantly higher than voice messages without speejis. One participant stated that \emph{``Initially my reflection was this introduces like an extra sort of layer of complexity. but then it makes it more efficient.''}, continuing with examples of how difficult it is sometimes to convey the tone of the message including the example of  when they write their partner \emph{``I am on my way home (hot emoji)''} means something completely different or when their father sends them a message stating "call me back" that that can sound like a threat to someone who does not know their father. Another participant highlighted the fact that voice messages are used when things are more serious and whenever they get a voice message they are extremely anxious about the content, having a smiley attached to the message gives them a lot of comfort, especially when you are not in a situation where you can listen to the voice message right away. These comments indicate that speejis are also associated with functional qualities and potentially the efficiency dimension measured by the UEQ questionnaire.

\subsubsection{What did participants not like?}
Most of the comments referring to what participants did not like about speejis focused on the added complexity, mainly associated with the coloured waveform  design but also concerns related to the fact that the augmentation was happening fully automatically, which is reflected for example by the following statement of one of our participants \emph{``it is a little hypothetical, but just the idea that if it got your intention wrong and maybe it was like a really sad message, but you were like, I don't know, nervously laughing throughout and it misunderstood that, it could create potentially uncomfortable situations.''}. Consequently, multiple participants stated editorial power would be good in any case, such as removing emojis or adding one more emoji to the existing was mentioned as examples. To clarify, none of the participants stated that the created emojis were useless or wrong. One participants stated, if the option (to select manually) is not default they probably would not bother to search for it in the settings but use whatever the default is. We also asked if marking the emojis as AI generated would help limit their concern and participants agreed that this idea would indeed remove their concerns. Consequently, an important take away is to make clear who added the emojis (user or AI). 
It is worth mentioning that some of the participants that mentioned some concerns about the coloured waveform  would however be more interested in receiving such detailed analysis as the receiver of voice messages.  When we asked why, one participant explained that it is more important to them to better understand the tones in messages others send to them then the other way around. When stated that this processing could happen automatically on the receiving side and how they would feel about this, there was a consensus that the sender should be in control if possible, but there was also some reflection on the fact that such things were out of their control and if it happened on the receiving device automatically it would also be fine. However, two participants highlighted while they would use this features in personal messaging they would not use it in some social network apps.  

\subsubsection{Self-reflection associated with the use of speejis}
Two participants stated some disagreement with the recognition results and during the interview we asked them to try it again and demonstrate what they mean. We played the audio back to go through the recognition in detail. With one participant we realized together that they were going down with their tone at the end of sentences and the message. They stated after realizing this  \emph{"I think it makes sense that for almost all of that it is interpreting me as being sad at the end, but that is not how, hmm, that is not my feeling."} and that this is how they speak. The other person had a native British accent, which was somewhat challenging for the text transcription they also spoke very fast. Anyway, when we asked them to demonstrate what they mean they were surprised themselves that \emph{``To be fair it actually did match my intend. I tried to start off happy and then neutral, sad and then happy at the end.''}, they explained then that the issue must have been with that there being no good emoji to match their neutral  speech. Overall, these conversations resulted in a discussion on how sometimes it may be better to not use what the AI correctly recognizes but allow the user to define the intend. There is indeed some justified concern that the AI emphasizing or explicating an emotional context that is there but not as strong as with the augmentations. On the other hand the augmentations may help reflect and realize how one sounds to others.   

\subsubsection{Emergent themes from the interviews}
To structure the analysis of the interviews, especially the open ended questions and topics brought in by the participants we followed the basics of thematic analysis~\cite{Braun01012006} and protocol analysis~\cite{ericsson1980verbal}, which included the creation of mind maps for each session and merging them iteratively to a few key descriptors and connections summarizing the the results  with respect to the subjective research objectives of us as researcher, which were understanding the impact of using speech emotion technology to augment and improve the UX of voice messaging.
The following, three themes provide an umbrella for the content of the interviews:
%three themes describe cross-cutting patterns of meaning from the interview data:

\begin{itemize}
  \item \textbf{Challenges for design},  a theme that summarizes all the comments related to the complexity of the  challenge to visualize speech emotion information as part of the voice message user interfaces.  Most of the concerns related to getting the coloured waveform design right, with suggestions on reducing the number of bars, focussing on only on parts of the waveform where the interesting things are happening, enabling a smoother transitions. Some comments also related to the use of emojis, including the number of emojis if neutral emojis are needed, if all parts of the message be used to create emojis.  Essentially a sequence of emojis seemed desirable especially for long message but the coloured waveform needed to work well with the emojis and the waveform designed needed to be less jumpy and less detailed. Suggestions also included making situated use of the linguistic parts of the message for the augmentations and exploring animations and interactivity to further expand expressiveness.  
  
  \item \textbf{Emergence of a new media}, is a theme that was more latent reflected in participants' awareness that AI is part of the communication  creating new form of communication which is inherently different and will impact how we communicate and our conversations. One participant mentioned the new way of communication with he following words, \emph{``I think it really touches upon something that is very deeply rooted in human conversation ... and I would really like to see it play out in a conversation. I'm still very sceptical, because it is then you have this third party [the AI] that comes with this.''}. The participant goes on reflecting why an additional party interpreting emotions could be both good and bad, providing the example of couple therapy, and arguing that a somewhat impartial interpreter of emotion can be useful but that there is also utility in emotions being ambiguous. This theme somewhat highlights participants concerns and predictions about a culture of communication that has not formed and developed yet but with such technologies will develop in foreseeable future. 
  
  \item \textbf{Recalibrations}, is a theme that we use to summarize all the self-reflective comments of the participants negotiating their new role where editorial power, power of emotion interpretation, control over data, etc.  is somewhat split between AI and the two humans who are part of an augmented  conversation, where the humans expose more (or potentially too much) of themselves and see more of the other party for the benefit of ``improved'' in terms of fun, and fast communication. However, there is also a risk that emotions are interpreted as something clear and somewhat deterministic while in human communication emotions are also useful when their interpretation is very ambiguous. One participant put it as \emph{``When we communicate, sometimes ambiguity is something that we want to avoid. And sometimes I think it's a good thing.''}. 
  
\end{itemize}

\section{Implications}

\begin{figure}[ht!]
  \centering
  \includegraphics[width=\linewidth]{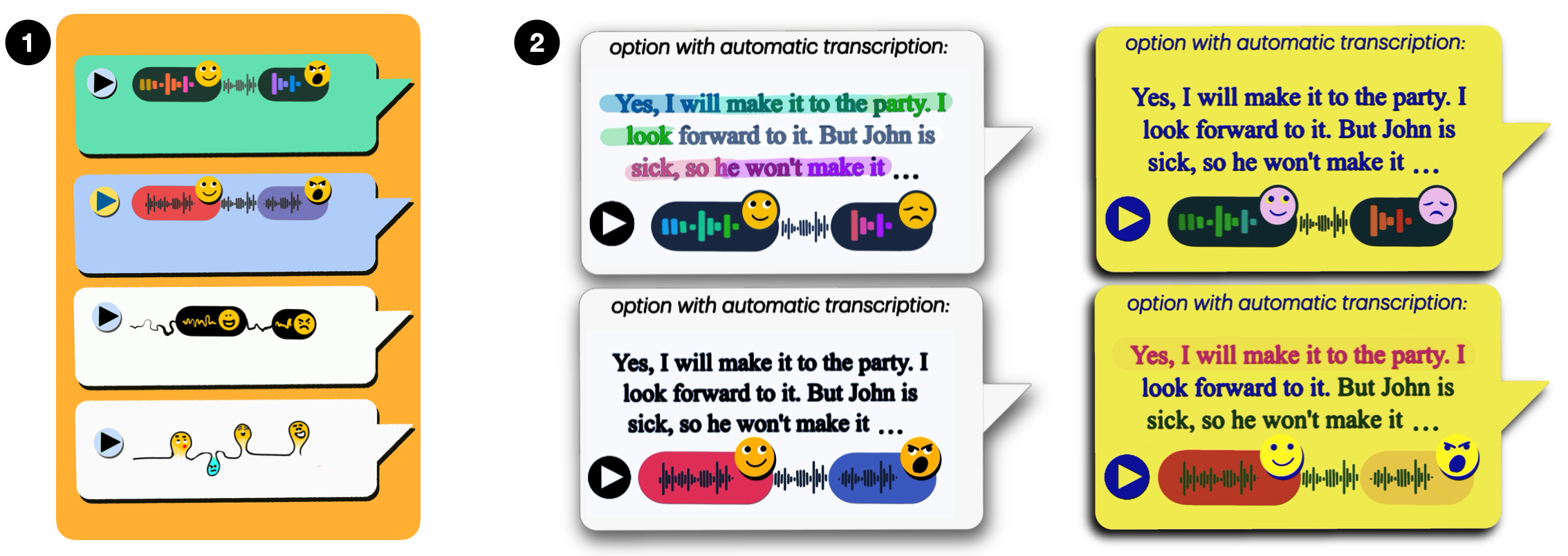}
  \caption{Iterated speeji design examples, addressing the issue with better connecting emojis with the waveform and dealing with longer messages while keeping the overall design playful and fun.}
  \Description{}
  \label{fig:design_iterations}
\end{figure}

Based on the results of the study, we performed a first round of iterations of the speejis design which are presented in Figure~\ref{fig:design_iterations}, \textbf{1} presents ideas to better combine emojis with the waveform design and use emoji sequences which would also be useful in longer messages or to describe the ``emotional ambiguity or richness'' that is inherent in human communication. The designs also skip parts of the message that would map typically to neutral emojis. We also try to reduce the complexity in the waveform by highlighting and colouring only the parts that are of interest (in terms of emotional expressions) while sticking to a joyful and fun design. Furthermore, we imagine that the emoji and the coloured waveform can work like an additional play button, where the receiver can play the affective parts of the message separately. 
Part \textbf{2} in \ref{fig:design_iterations} shows some examples of how the speejis work with transcriptions where approximate connections between text and augmentations could be created and through highlighting text parts or using mapping colours for the font. With the emojis being attached to the audio waveform and not part of the text, the designs somewhat differentiate themselves as being AI generated emojis and not text input from the user.

\begin{figure}[ht!]
  \centering
  \includegraphics[width=\linewidth]{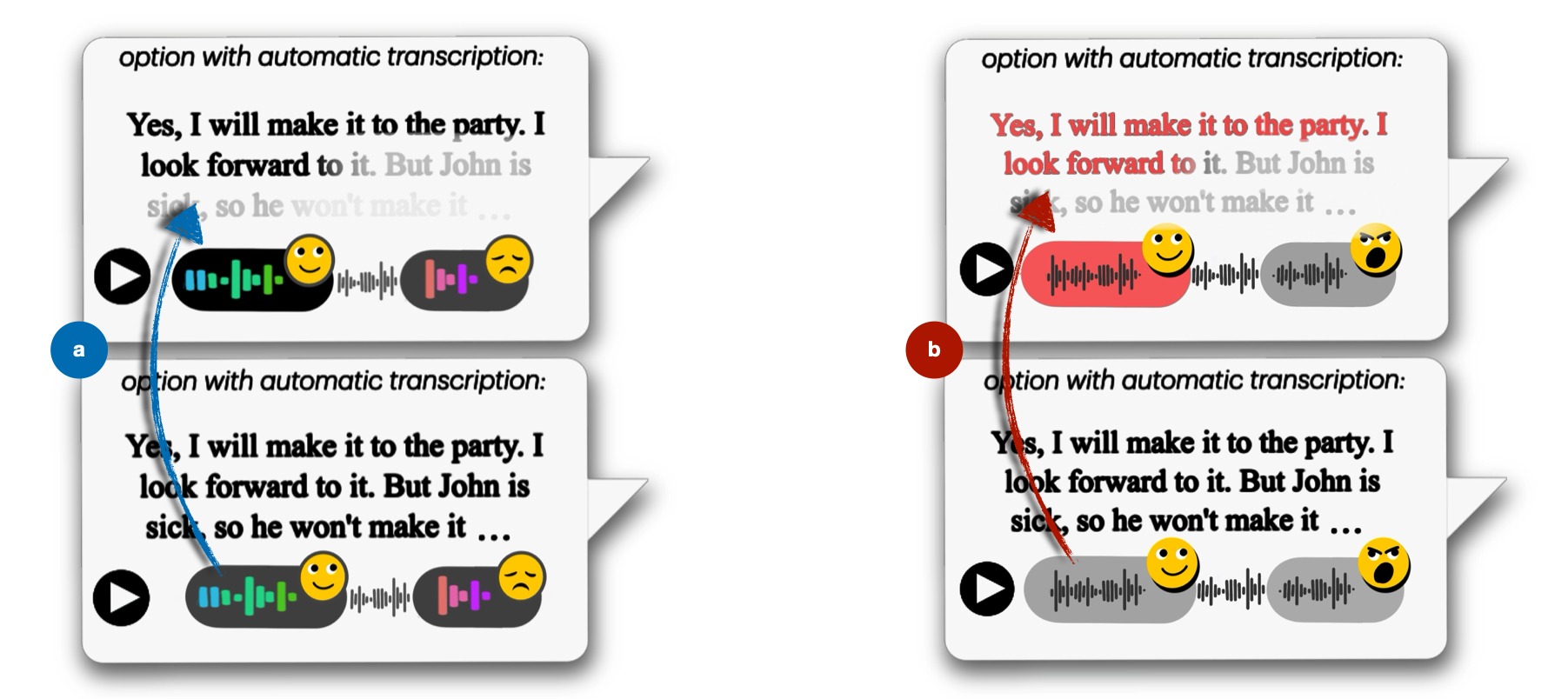}
  \caption{ Iterated speeji design examples, aiming to reduce  the issue of complexity with especially long messages by allowing to tab a speeji, and thereby enabling users to skip, play, or transcribe and read the part of the voice message that the speeji provides locus to.}
  \Description{}
  \label{fig:design_iterations_complexity}
\end{figure}

Figure \ref{fig:design_iterations_complexity} shows further design ideas to reduce complexity through interactivity, especially with longer text, introducing some interactivity where pressing a speeji may cause the transcription or highlighting of that part. The option to transcript only the part of the message based on user action would have the added benefit of being somewhat more sustainable, since in our experiment the transcription model required more time and some delay (e.g. 2-3 seconds) to transcribe longer text sequences compared to SER model which provided results without a delay. We could also imagine that a dedicated emoji and colour could be used to augment the whole bubble to provide some overall tone.

With the implications for design we mainly address the ``challenges for design'' theme we summarized in in the results section. 
Our design iterations a examples and inspirations for future directions but are not to be regarded the only design direction.
For example, using short non-verbal audio clips could be an additional way to indicate the tone of the voice message. In this context one participant suggested to use the voice message as a starting point to create  short non-verbal sound. Previous research studying the personalization of emotional sounds for social robots has used a similar task as a starting point, cat sounds~\cite{mertes_et_al_2019}. 
As a matter of fact, personalization or adaptation is an important future direction and something that is required for emotion communication, addressing the participants insights summarized in the ``calibrations'' theme. Augmentations should ideally adapt to a specific persons edits. For example, when a person always sounds sad for the AI but the person corrects this augmentation than the augmentation model should personalize towards this style of the user and not augment the expressions as sad in future and thereby enabling the user to own how their emotional expressions should be 
interpreted.

Studying emotions and emotional communication is challenging for multiple reasons, including that our understanding as a discipline or in sciences in general for what emotions are in human-human communication being very limited, let alone when we add a machine or AI as a third party that can sway the emotion interpretation and conversation towards a direction that may be wrong or to one-sided. Most of the methodology with regards to UX that we have is targeted towards human-computer or human-product interaction. It would be desirable to have better or new methodologies that can help understand the impact and role of third-party AI and their augmentation in human-to-X interactions especially when the topic is about emotions, which are deeply rooted in human relationships.

\section{Discussion}

The design of mobile user interfaces is changing with ever more elements using the help of AI to design and automate user interfaces that are more proactive and co-creative. Previously, mobiles have helped computing become ubiquitous, allowing users to digitally communicate and work in different contexts while staying socially connected to others. The joy of using mobiles is also linked to the ways they allow us to create media and share our experiences in many ways with friends and family. In this context the process of AI, especially the so called deep learning era had a large impact on new possibilities to capture, modify, and create content.

In this paper, we focused on novel affective computing functions, which are typically used to recognize and stimulate user emotions and affective behaviour. One type of interaction designs that benefit from mobiles becoming more perceptive and expressive of affect are implicit interactions. Mobiles have always benefitted from more context information to deliver customized experiences based on location, language, and user profiles; often based on explicit information provided to the mobile services and devices. Doing this enabled mobiles to address limitations of usability on small sized screens and keyboards and contextual accessibility.  

Today, mobiles are increasingly capable of recognizing nuances of user behaviour implicitly, which can help mobiles to become more efficient in being helpful by predicting user intents before they are explicitly communicated. Of course, this type of proactive machine involvement has to be treated cautiously, as we have also seen in our user study, users prefer transparency and very careful  attribution and accreditation of AI generated content and behaviour.  
On the other hand, AI developments are making new functions and new forms of communication possible. As we describe in this work with the use of speech emotions processing to create speejis, automatically added visual cues to augment voice messages. 
We showed that doing this, not only significantly impacts the user experience, but provided added value and contextual accessibility for the price of being more dependable on what the AI does. This is also a main difference our work exposes and discloses compared to related and previous work. 

We are not the first to recognize benefits of using speech emotion recognition technology, but to the best of our knowledge, this work is the first to use speech emotion recognition to produce voice message augmentation.
For example, the EmoWear system~\cite{an2024emowear} proposed using SER to reduce the number of choice for voice augmentation from six emotion categories to two and from 30 bubble animations to ten. 
Their evaluation thus focuses mainly on the effect of manually selecting, sending,  and receiving animated voice message teasers on a smartwatch and not on the effect and utility of SER in this process. In comparison we have focused on the performance of the AI taking over this task on behalf of the user considering that there are many situations where due to contextual accessibility and convenience reasons this action could be done by an AI. 

This is quite different from a large body of exiting work in augmenting messaging (e.g.,~\cite{Haas_2020_mobileHCI, an2025aniballoons}), where the focus is on creating new augmentations visual and auditive for users to edit and select manually. The limitations we have in return is that we did not spend as much focus on the detailed design of speejis but instead discovered valuable knowledge to improve the design of speejis and an understanding of what we gain (e.g., stimulation, novelty, and attractiveness) and potentially lose (e.g., dependability). 

Participants' mentioning of animations as an option which would add to the liveliness of the emotional design would be a form of social and proxemic behaviour and in line with previous work in UX design suggestions a paradigm shift moving from the designing of static GUI elements to elements that implement dynamic proxemic behaviours~\cite{Aslan_Andre_2017}. We have also come to understand that speejis as a form of AI driven communication augmentation has a transformative quality, which we also saw in participants comments about speejis impacting conversations, e.g.,  similar to how AI generated text summarizations can impact human-human communication. 

We argue that the style of communication we studied can be seen as (part) of a new media form, which is emerging. This is where we can take lessons from past work theorizing about media usage and the emergence of new media, especially the work of McLuhan~\cite{mcluhan1967medium} who puts emphasis on how new media is not only a medium that carries the message but is a message in itself. What is then the message of AI augmenting human voice messages with speech emotion interpretations one wonders?
Maybe the fact that in future messaging is an increasingly co-creative act. We can expect that such new media will likely change conversation in unpredictable ways, our study showed some indications on how communication may change. In the positive sense, we will see some improvement in accessibility, usefulness, and efficiencies in getting the intended message across and quickly receiving the feeling of the message. On the negative, we are moving away from a user-centred design towards designs that address some symbiotic relationship with between a user and an ``AI'', where we have to address both the pair of these two and the needs of the individual user in this relationship with our research and design studies. 
To this end and as we indicated already in the implications, new methodologies will be needed rapidly to keep up with how AI is being already applied and used to shape user experiences. 

\subsection{Limitations and Outlook}
As we are entering unexplored new fields with AI-based solutions taking a proactive and experience-shaping role, our work faces multiple limitations. Some of these limitations relate to a lack of methodology and tools to especially evaluate proactive interactions. In our case the ``AI'' proactively augments one's voice messages with speech emotion cues including emojis, this proactive behaviour is under-explored in HCI research in contrast to interaction with reactive systems. 
Some of the limitations are more practical, such as the emoji mapping by Kutsuzawa et al.~\cite{kutsuzawa2022classification} not providing any labels for the dominance dimensions while the SER model by Wagner et al.~\cite{wagner2023dawn} provides labels for the dominance dimension. We are aware that dominance is a dimension that could be useful in augmenting conversations in future since it is an emotion dimension that is relational and apparent in social interactions. We also face some limitations in terms of juxtaposing emojis and emotional waveform designs in the UEQ ratings, which makes it harder to understand how much which type of speejis impacted the ratings in positive or negatives way. Although, we compensated with reporting on participants preferences and discussing the topic in the interviews and reflecting on them in the analysis of the interviews. 

A main limitation and something we from the beginning did not plan to address is testing this functionality in the wild and as part of real usage in chat conversations, which is something that we have to look into in our future work. However, we would predict that implementing this functionality into a real messaging application will come with additional technical challenges which are mainly relevant to the solution being deployable, such as the use of speech enhancement  and voice activity detection models to optimize robustness for use in the wild, which of course goes beyond the scope of this first study into this specific topic understanding the UX and user needs of speejis.   

While the UEQ measures show no significant difference for the efficiency dimension, we would argue that efficiency benefits were not clearly experienced considering the way we setup the study. In our study participants were encouraged to explore how voice messaging feels but they were not asked to complete a task under time pressure or put into a situation were efficiency was clearly experiences. In contrast, the participants when encouraged to reflect more, including reflection on the benefits of both sending and receiving such messages appreciated the fact that it was not only fun to use but that one could get much quicker and easier a sense of what is going on. There we have a clear limitation in our initial study setup, which while providing very useful insights and results considering look and feel, lacked deeper results in completing tasks and goals. Another limitation we want to mention is that we did separately measure how much of the fun part etc. was linked with using emojis and how much the effects were to attribute to the coloured waveform. We have some indications that are reflected in participants stating that waveform and emojis worked well together and none of the participant stated to prefer a design without a waveform. But clearly there were some issues associated with the waveform and less with using emojis. This dilemma combined with the fact that interpretations of emojis can be somewhat subjective could be reflected in the lower results for the dependability dimension of the UEQ. 

New emoji sets specifically designed to be used as speejis are desirable to optimize the experience, such as the efforts we see in related work by An et al.~\cite{an2025aniballoons} creating and mapping chat-balloon animations to continuous emotion values. Such mapping could enable automatic mappings of speech emotion recognition results with dedicated visual speeji designs. Previous work has also suggested the potential to augment voice messages with audio effects~\cite{Haas_2020_mobileHCI}, manually applying filters and background sounds to enrich media experiences. Thus, future speejis may not be restricted to the design of visual cues but also encompass audio effects or even haptic feedback to be used e.g. as teasers for an incoming voice message.

\section{Conclusion}
This paper addressed the topic of using speech emotion recognition technology to implement a new proactive way to augment voice messages with visual speech emotion cues, such as emojis. We presented a first prototype implementation and conducted a user study demonstrating the impact on UX that speejis has compared to a baseline. We provide concrete design implications and insights into user needs and concerns of having an AI interpret one's emotions and help augmenting voice messages to achieve better messaging accessibility and experiences. Our results demonstrate the potential of speejis on many UX constructs in messaging, though at the cost of some dependability on AI that is acting proactively, taking initiative, and co-creating the voice messaging experience. We highlighted both opportunities and concerns for future research in this context.

%\section{Acknowledgments}
%\section{Appendices}

%%
%% The acknowledgments section is defined using the "acks" environment
%% (and NOT an unnumbered section). This ensures the proper
%% identification of the section in the article metadata, and the
%% consistent spelling of the heading.
%\begin{acks}
%To Robert, for the bagels and explaining CMYK and color spaces.
%\end{acks}

%%
%% The next two lines define the bibliography style to be used, and
%% the bibliography file.
\bibliographystyle{ACM-Reference-Format}
\bibliography{References}

\end{document}